\documentclass[trackchanges]{aastex701}

\begin{document}

\title{High Resolution Optical Methane Linelist from observations of Titan for Cross-Correlation studies}

\author[orcid=0000-0002-0586-9536,sname='Rianço-Silva']{Rafael Rianço-Silva}
\affiliation{Department of Physics and Astronomy, University College London, Gower Street, WC1E 6BT London, United Kingdom}
\affiliation{Instituto de Astrofísica e Ciências do Espaço, Universidade de Lisboa, OAL, Edifício Leste, Tapada da Ajuda, PT1349-018 Lisbon, Portugal}
\affiliation{Departamento de Física, Faculdade de Ciências, Universidade de Lisboa, Edifício C8, Campo Grande, PT1749-016 Lisbon, Portugal}
\email[show]{rafael.silva@ucl.ac.uk}  

\author[orcid=0000-0001-6757-5763, sname='Machado']{Pedro Machado} 
\affiliation{Instituto de Astrofísica e Ciências do Espaço, Universidade de Lisboa, OAL, Edifício Leste, Tapada da Ajuda, PT1349-018 Lisbon, Portugal}
\affiliation{Departamento de Física, Faculdade de Ciências, Universidade de Lisboa, Edifício C8, Campo Grande, PT1749-016 Lisbon, Portugal}
\email{pmmachado@ciencias.ulisboa.pt}

\author[orcid=0000-0002-7853-6871, sname='Sousa Silva']{Clara Sousa Silva} 
\affiliation{Instituto de Astrofísica e Ciências do Espaço, Universidade de Lisboa, OAL, Edifício Leste, Tapada da Ajuda, PT1349-018 Lisbon, Portugal}
\affiliation{Bard College, 30 Campus Road, Annandale-on-Hudson, NY 12504, USA}
\email{csousasilva@bard.edu}

\author[orcid=0000-0001-9286-9501, sname='Yurchenko']{Sergey Yurchenko} 
\affiliation{Department of Physics and Astronomy, University College London, Gower Street, WC1E 6BT London, United Kingdom}
\email{s.yurchenko@ucl.ac.uk}

\author[orcid=0000-0001-6058-6654, sname='Tinetti']{Giovanna Tinetti} 
\affiliation{Department of Physics and Astronomy, University College London, Gower Street, WC1E 6BT London, United Kingdom}
\affiliation{NMES Faculty, King’s College London, Strand Building, Strand, WC2R 2LS London, United Kingdom}
\email{giovanna.tinetti@kcl.ac.uk}

\begin{abstract}

Exoplanet atmosphere characterization heavily relies on molecular spectroscopic data. Despite efforts to obtain comprehensive spectral libraries for the chemical characterization of exoplanet atmospheres, large gaps remain, particularly for larger molecules and higher frequencies at high spectral resolution. One key example is the methane (CH$_4$) optical spectrum. CH$_4$, the simplest hydrocarbon, is a crucial species for exoplanet atmosphere characterization and a possible biosignature. However, until now, high-resolution linelists at optical wavelengths for CH$_4$ have been very challenging to obtain either experimentally or computationally, leaving the high resolution spectrum of CH$_4$ uncharacterised across most of the visible spectrum. This restricts exploration of CH$_4$ absorption in the optical regime, as upcoming instruments such as ELT-ANDES and VLT-RISTRETTO will start probing the atmospheres of ever smaller exoplanets in optical wavelengths. To address this spectroscopic data limitation, we observed Titan’s optical spectrum – dominated by CH$_4$ absorption – at the highest spectral resolution to date with VLT-ESPRESSO. From it, we produced an empirical, low-temperature high-resolution (R $\sim$ 190000) linelist of CH$_4$ in optical wavelengths which we present here, with thousands of previously unidentified lines. We employ this CH$_4$ linelist (RRS-2026) to build a template suitable for high-resolution cross-correlation spectroscopy (HRCCS) studies – a first for CH$_4$ in optical wavelengths. With this new linelist, we performed the first HRCCS detection of CH$_4$ in the atmospheres of Titan and Jupiter using optical high resolution spectra. This work sets the stage for the search for CH$_4$ in exoplanet atmospheres through HRCCS with current and future ground-based high-resolution optical spectrographs, showcasing how Solar System observations provide useful products for exoplanet research.

\end{abstract}



%%%%%%%%%%%%%%%%% BODY OF PAPER %%%%%%%%%%%%%%%%%%

\section{Introduction}
Understanding the physical and chemical properties of planetary atmospheres, both within and beyond the Solar System, is central to modern planetary science and exoplanet research. Over the past decades, this effort has been propelled by rapid advances in remote-sensing techniques, sustained by parallel advances in our understanding of molecular spectroscopy \citep{Chubb2024,Tennyson2016, gordon2026hitran2024}. These studies have enabled an in-depth characterization of the bulk compositions and minor constituents of all major Solar System atmospheres \citep{Encrenaz2015}, while also allowing a preliminary survey of the different atmospheric compositions across exoplanet populations with state-of-the-art space telescopes such as JWST \citep{Yurchenko2025}. 
\par The dawn of High Resolution Cross-Correlation spectroscopy (HRCCS) \citep{Snellen2010,Birkby2018,Snellen2025} has taken these remote sensing studies beyond their previous limits, using high-resolution transmission spectra of exoplanets ($R \equiv \frac{\lambda}{\Delta \lambda} >$ 25 000) to combine multiple weak spectral lines (too weak to be individually identified) into a single observable, a Cross-Correlation Function (CCF), with increased SNR (Signal to Noise Ratio), allowing previously impossible molecular detections, and providing an unprecedented observable of exoplanet atmosphere dynamics from the Doppler signals on the CCF \citep{Birkby2018,Snellen2025}. This is done through cross-correlation of the observed transmission spectrum with a spectrum template of the molecule of interest, on which the effectiveness of the CCF detection is highly dependent \citep{Yurchenko2025,Brogi2019,Gandhi2020}.

\par A parallel improvement on the quality, spectral range, spectral resolution, and temperature range of the available molecular spectral linelists has been taking place, through initiatives such as ExoMol \citep{Tennyson2012}, HITRAN \citep{HITRAN}, or HITEMP \citep{HITEMP}. However, despite these efforts, important gaps in our available spectral libraries still remain, particularly for larger, more complex molecules, as well as higher frequency regimes and spectral resolutions \citep{Chubb2024,Yurchenko2025}. One molecule for which this is particularly evident is methane (CH$_4$), for which neither of the aforementioned spectral libraries present CH$_4$ linelists in visible wavelengths ($\lambda < $ 700 nm) \citep{Yurchenko2024,Hargreaves2020,Bertin2024}.
The challenges posed by the extraction of CH$_4$ visible spectrum linelists are felt both in theoretical and laboratory studies \citep{Bourdon2009,Campargue2023}. As the number of energy sub-levels and possible transitions increases at higher energy polyads, CH$_4$ near-infrared (and especially) visible polyads can contain up to millions of weak transitions \citep{Bourdon2009}, very challenging to model through quantum \textit{ab initio} simulations \citep{Yurchenko2024,Kefala2024}, but also too weak to be detected in laboratory studies, even through considerably long pathlengths \citep{Campargue2023}. Hence, our knowledge of CH$_4$ spectrum at $\lambda <$ 700 nm is mostly supported by low-resolution empirical cross-sections \citep{Giver1978,Karkoschka1994,Karkoschka2010}, most of them obtained from observations of Solar System gas giants or Titan, where long atmospheric pathlengths and abundant CH$_4$ cause weak optical absorption bands to stand out in their atmospheric spectra \citep{Bourdon2009,Karkoschka2010}.

\par The lack of spectral knowledge regarding the visible high-resolution spectrum of CH$_4$ causes its detection through HRCCS to be possible only in infrared wavelengths \citep{Yurchenko2025} where HRCCS detections are more challenging due to the increased opacity of Earth's atmosphere \citep{Birkby2018}. Thus, extending high-resolution linelists of CH$_4$ to visible wavelengths could enable a complementary search for this compound in exoplanet atmospheres, with instruments such as VLT-ESPRESSO \citep{Pepe2014} or upcoming instruments such as VLT-RISTRETTO \citep{Lovis2022} or ELT-ANDES \citep{Palle2025}. Indeed, the goal of retrieving high-resolution reflection spectra of habitable-zone exoplanets in visible wavelengths to search for possible biosignatures, such as CH$_4$ \citep{Lovis2022,Martins2016}, requires CH$_4$ high-resolution linelists in visible wavelengths.

\par CH$_4$ is a considerably important chemical species for the study of planetary atmospheres. In warm and cold giant exoplanets, CH$_4$ is thought to be one of the dominant volatile species \citep{Fortney2020} as it is for giant planets in the Solar System  \citep{Karkoschka1998,Irwin2019}. It has also been identified as a major atmospheric constituent in many volatile-rich sub-Neptunes \citep{Madhusudhan2025} and Solar System icy moons like Titan \citep{Karkoschka2010} where it is known to shape its climate \citep{Lunine2008} and serve as the molecular source of Titan's abundant and complex photochemistry \citep{Hayes2018}. The presence of CH$_4$ in telluric planets is often regarded as a marker of chemical disequilibrium, potentially as a biosignature \citep{Thompson2022}.

\par Given its great scientific importance, recent works have aimed to obtain observation-based high-resolution CH$_4$ spectra. In 2024, high-resolution ($R \simeq$ 100 000) observations of Titan with VLT-UVES yielded 97 previously undetected high-resolution spectral lines on Titan's spectrum between 540 nm and 620 nm \citep{rianco-Silva2024}. In 2025 another study retrieved an ``intrinsic spectrum of Titan'', presenting the retrieval of up to 6195 spectral features that could be associated with CH$_4$ on a VLT-ESPRESSO ($R \simeq$ 190 000) spectrum of Titan \citep{Sithajan2025}, though we find that some of the detected spectral features may be affected by the presence of overlapping solar spectral lines which undermine its applicability to HRCCS studies \citep{Yurchenko2025} (see Discussion section).

\par Hence, with the explicit goal of retrieving a high-resolution linelist (containing line positions and relative line intensities) that could be used for a HRCCS visible CH$_4$ template, in this study we obtained a new VLT-ESPRESSO UHR spectrum of Titan with an increased SNR, in 2024. We use it alongside the 2021 VLT-ESPRESSO Titan spectrum used in \cite{Sithajan2025}, and a telluric calibrator star also observed with VLT-ESPRESSO, to extract an accurate though conservative low-temperature linelist of CH$_4$ from Titan's atmosphere. We then build a CCF template of CH$_4$ in optical wavelengths and for the first time perform a HRCCS detection of CH$_4$ in a planetary atmosphere in optical wavelengths on VLT-ESPRESSO observations of Jupiter. Finally, we compare our linelist with past CH$_4$ optical spectra - particularly with low-resolution cross-sections which for a long time had been our sole glimpse into the CH$_4$ visible spectrum.

\section{Observations and Data Reduction}

The observations employed in this study were carried out with the ESPRESSO spectrograph (Échelle SPectrograph for Rocky Exoplanets and Stable Spectroscopic Observations) mounted on ESO’s Very Large Telescope (VLT). ESPRESSO is a fiber-fed, cross-dispersed échelle instrument operating at visible wavelengths in the Incoherent Combined-Coudé Laboratory (ICCL) of the VLT, where it can be supplied with light from any of the four 8.2 m Unit Telescopes \citep{Pepe2014}. While the instrument was primarily designed for the detection and characterization of exoplanets, ESPRESSO has also demonstrated its usefulness in solar system research, including Doppler velocimetry of Jupiter’s atmosphere \citep{Machado2023}.

\par Titan was observed for a total duration of 2 h 20 min with ESPRESSO at the VLT, beginning at 00:17 UTC on 4 December 2024, under ESO program 114.277N (PI: P. Machado), as described in \cite{Rianco-Silva2025}. The observations were conducted in the Ultra-High-Resolution (UHR) mode \citep{Pepe2021}, providing the instrument’s maximum resolving power of $R \sim 190000$, delivering the highest spectral resolution dataset of Titan obtained to date in visible wavelengths. Atmospheric conditions were excellent throughout the run, with clear skies and a seeing below 0.8 arcsec (FWHM). The 0.5 arcsec fiber was completely illuminated by Titan’s apparent disk (0.83 arcsec across). At the time of observation Titan had a V-band apparent magnitude of +8.62, a surface brightness of 7.71 mag arcsec$^{-2}$, and was 99.73\% illuminated. The relative velocity of Titan with respect to Earth was 24.4 km/s and Titan's velocity with respect to the Sun was -5.3 km/s at the moment of the observations - according to the JPL Horizons ephemeris calculator \citep{Horiz}. Seven individual exposures, each 20 minutes long, were obtained, yielding complete spectral coverage of Titan's backscattered spectrum across the 378.2–788.7 nm range. Each individual exposure reached SNR above 350 at 500 nm \citep{Rianco-Silva2025}.

\par Beyond Titan, the 114.277N observing program on the night of 4 December 2024 also observed the V=6.43 magnitude A0V-type star HD 218639. This yielded 3 exposures (240 s each) of this star, each with a typical SNR of 200 at 500 nm - which were useful to calibrate for Telluric Absorption lines. We shall refer to these datasets as the ``Titan 2024'' and ``Star 2024'' datasets.

\par Our dedicated observations of Titan with VLT-ESPRESSO for this study are complemented by older, archived, publicly available VLT-ESPRESSO observations of Titan, obtained in July 21, 2021 by the 106.218L program (PI: M.Turbet). These observations resulted from a shorter observing time of Titan (3 exposures of 960s each) at the same UHR mode we used, with the 0.5 arcsec fiber also entirely illuminated by Titan, yielding a total SNR of 280 at 500 nm, after stacking all observations. We shall refer to this dataset as ``Titan 2021''.

\par Finally, for an independent comparison of the retrieved CH$_4$ linelist, we used archived VLT-ESPRESSO observations of Jupiter (which also contains CH$_4$, albeit in a drastically different environment). This spectra was also obtained at the UHR mode of ESPRESSO, on the night of July 21, 2019 by the 0103.C-0203(A) program (PI: P.Machado, published in \cite{Machado2023}). Jupiter's surface brightness was 5.44mag/arcsec$^2$, the ESPRESSO fiber 0.5 arcsec was placed in a position corresponding to a latitude of 10ºS and a longitude of 5,24º West from the
sub-solar meridian within the 43.8 arcsec diameter planetary disk. The 60s exposure yielded a SNR of 480 at 500 nm. We refer to this dataset as ``Jupiter 2019''.

\par All spectral observations were processed using the ESPRESSO Data Reduction Software version 3.3.0. After spectral flux calibration, the spectral continuum was fit for each exposure and normalized by its spectral ratio with respect to continuum of the exposure at the lowest airmass - following the approach used in \cite{rianco-Silva2024}. This enables a correction of the varying airmass's impact on the spectra across several exposures - allowing the exposures to be summed, to obtain a single spectrum for each of the 2 observation campaigns of Titan (Titan 2024 and Titan 2021), as well as a single spectra for the observation of the calibration star HD 218639 and for Jupiter 2019. The final spectral flux per wavelength bin for each of the summed spectra, $S_\lambda$,  was obtained as an average of the spectral fluxes for that wavelength bin across the $N$ summed exposures (eq. \ref{eq:average_flux}) whereas the spectral flux error per wavelength bin, $\delta  S_\lambda$, was obtained as the RMS sum of the spectral flux error for that wavelength bin for the summed exposures plus the scatter variance the exposure's spectral fluxes with respect to the average, $S_\lambda$, as shown in eq. \ref{eq:average_error}.
Each of these 4 VLT-ESPRESSO spectra (Titan 2024, Titan 2021, Star 2024, Jupiter 2019) were then normalized using a Savitzky-Golay filter \citep{Savitzky-Golay}, which effectively removed their spectral continuum, leaving behind only spectral absorption lines.
\begin{equation}
    S_\lambda = \frac{1}{N} \sum^N_i S_{\lambda, i}
    \label{eq:average_flux}
\end{equation}

\begin{equation}
    \delta S_\lambda = \sqrt{\frac{1}{N} \left(\frac{1}{N} \sum^N_{i=1} \delta S_{\lambda, i}^2 + s^2 \right)}, \text{ where } s^2 = \frac{1}{N-1}\sum^N_{i = 1}(S_{\lambda, i} - S_\lambda)^2
    \label{eq:average_error}
\end{equation}

\par One last spectrum used for this analysis is the high-resolution solar transmission spectrum obtained by \cite{Kurucz2006}. This ultra-high resolution spectrum covers the entirety of the VLT-ESPRESSO spectral range at resolving powers of $R =$500000. In order to be directly comparable to VLT-ESPRESSO UHR spectra at $R = 190 000$, we decreased the spectral resolution of the Kurucz+2006 solar spectrum \citep{Kurucz2006} to VLT-ESPRESSO UHR resolution by convolving it with a Gaussian curve with a FWHM equivalent to that of a spectral line seen at VLT-ESPRESSO's R = 190000 resolving power at $\lambda$ = 4000\AA, i.e., a FWHM = $\lambda/R$ = 0.021\AA. We picked the FWHM for $\lambda$ = 4000\AA, leading to a slightly higher spectral resolution (i.e., smaller FWHM) than ESPRESSO’s UHR at $\lambda >$ 4000\AA \hspace{0.1mm} in the solar spectrum we use to identify solar lines. This is a conservative approach, ensuring that no relevant solar spectral feature would be smoothened to below the ESPRESSO UHR spectral resolution at any wavelength.

\section{Results}
\subsection{Empirical CH$_4$ Optical Linelist extraction} The Titan visible spectra observed at high spectral resolutions contains a diverse set of features that must be understood before any CH$_4$ linelists can be extracted. The visible spectra of Titan is effectively shaped by four contributions: backscatter of visible solar radiation on Titan's atmosphere (reflecting the solar continuum and solar absorption lines \citep{rianco-Silva2024}); continuous Rayleigh scattering; and haze and CH$_4$ absorption on Titan's atmosphere \citep{Lorenz1999}, \citep{McKay2001}.

\par After normalizing Titan's spectra (effectively removing continuum contributions to the spectrum), we are left with a large set of spectral lines. These include absorption lines resulting from CH$_4$ absorption in Titan's atmosphere, which we are interested in uncovering - but also solar absorption lines (backscattered from Titan's atmosphere) as well as telluric absorption lines \citep{rianco-Silva2024}. It is thus key to efficiently distinguish between these spectral lines' origins when aiming to extract a CH$_4$ linelist from Titan's visible spectrum.

\begin{figure}%[tbhp]
\centering
\includegraphics[width=0.5\linewidth]{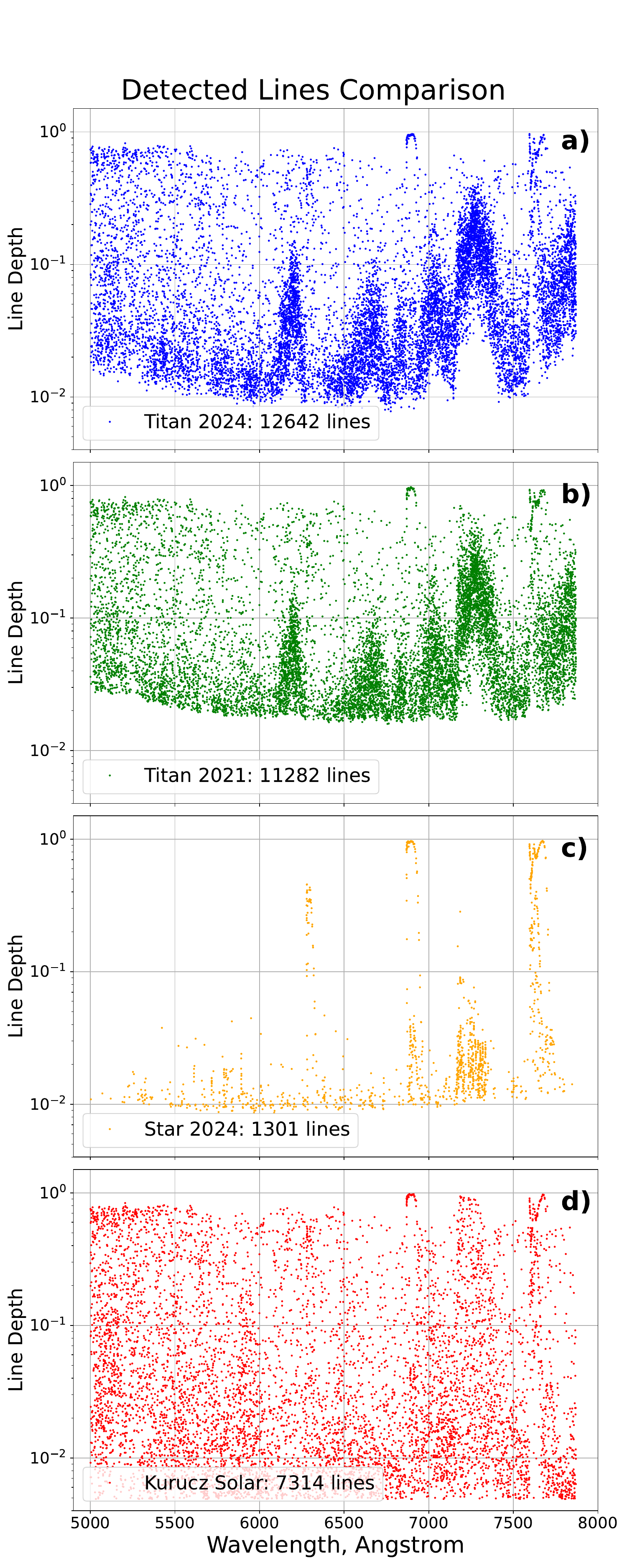}
\caption{Line detections above 5$\sigma$ of spectral noise for the spectra: a) Titan 2024, b) Titan 2021,  c) Star 2024 (corresponding to the telluric calibrator observation of star HD 218639) and d) Kurucz Solar.}
\label{fig:Detected_Lines_4_targets}
\end{figure}

\begin{figure}%[tbhp]
\centering
\includegraphics[width=0.85\linewidth]{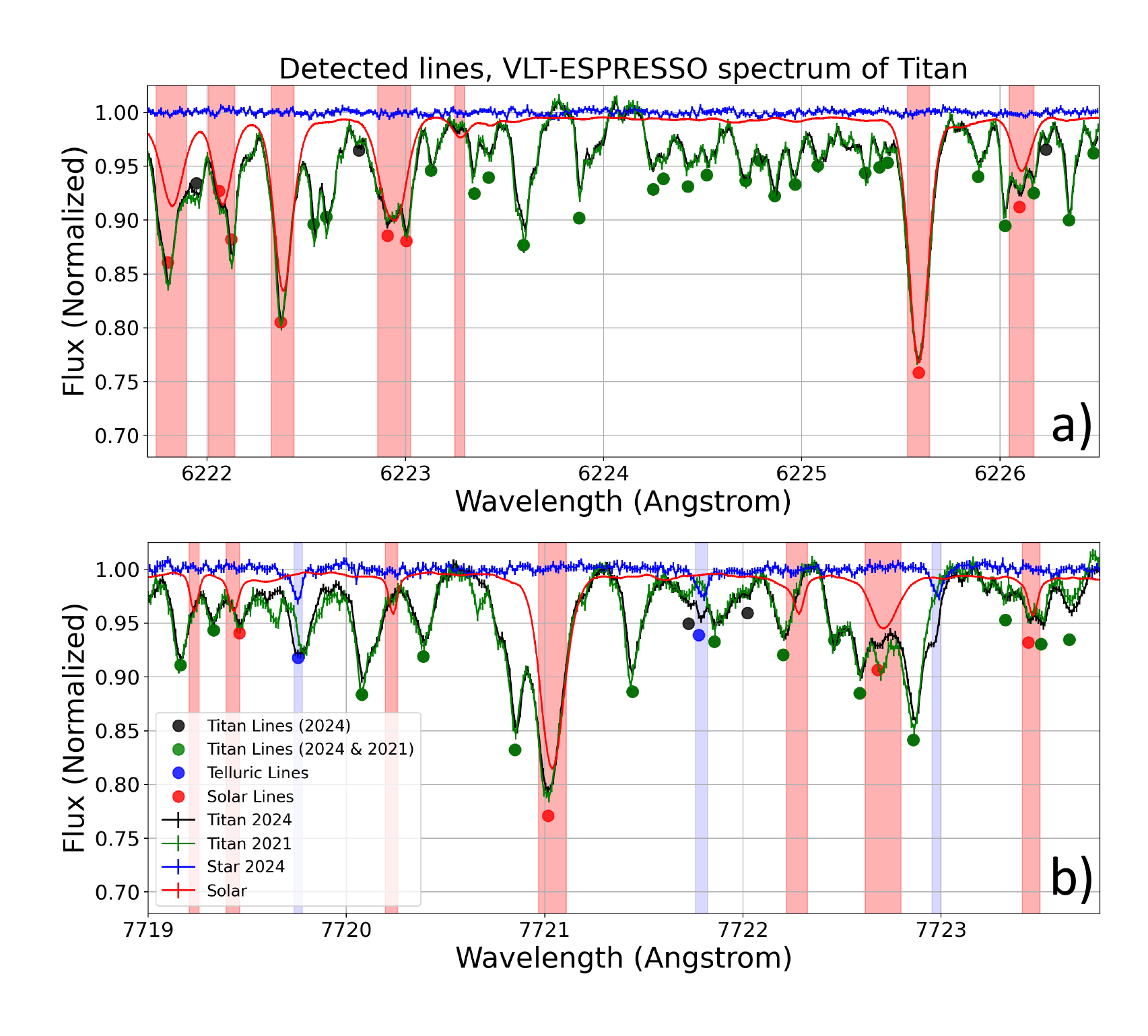}
\caption{Two examples of the spectral line characterization, comparing Titan 2024 (black), Titan 2021 (green), Solar (red), and Stellar telluric calibration (blue) spectra, and showcasing spectral line categorization as ``Titan'', ``Solar'' or ``Telluric'' in origin. Titan lines are split between the ones which are identified in both Titan 2021 and Titan 2024 spectra (green circles) and the ones which are only identified in Titan 2024 (black circles), due to the higher SNR of the observed 2024 Titan spectrum. Red circles identify lines in Titan spectra that are associated to Solar absorption features, and blue circles identify lines in Titan spectra associated to tellurics. Red-shadowed regions correspond to the measured FWHM of spectral lines in the solar spectrum, whereas blue-shadowed regions correspond to the measured FWHM of spectral lines in the ``Star 2024'', associated to telluric absorption lines.}
\label{fig:lines_spectra}
\end{figure}

\par Our approach is based on \cite{rianco-Silva2024}, where we take advantage of the identification of lines across distinct spectra (Titan, Calibration Star and Solar) to assess their origin. For this, we ran a line detection algorithm that identifies the absorption lines on Titan 2024, Titan 2021, Kurucz Solar and Star 2024 spectra (with line depths above a line depth threshold of 5$\sigma$, or 5 times the local spectral flux error, $\delta S_\lambda$), which are shown in figures \ref{fig:Detected_Lines_4_targets} (a to d). This yielded 12642 detected absorption lines on the Titan 2024 dataset, 11282 on the Titan 2021 dataset, 7314 on the Kurucz Solar dataset, and 1301 on the Star 2024 dataset.
For each detected line, we extract line wavelength, relative line depth and FWHM. Visual inspection of figures \ref{fig:Detected_Lines_4_targets}a and \ref{fig:Detected_Lines_4_targets}b shows a shared spectral structure in the detected lines on both Titan spectra (2024 and 2021). The more stringent lower limit of line depth detection on Titan 2021 results from its comparatively lower SNR compared to Titan 2024. This distribution of detected line depths vs wavelength is reminiscent of the visible cross-sections of CH$_4$ in visible wavelengths \citep{Giver1978, Karkoschka1994, Karkoschka2010} indicating that many of these newly detected spectral features correspond indeed to CH$_4$ absorption in Titan.
\par However, the large number of solar absorption features detected in the Kurucz Solar \citep{Kurucz2006} spectrum indicates that many of the detected features on Titan 2024 and Titan 2021 (scattered away from the denser line detection distribution that likely corresponds to CH$_4$ features) are of solar origin. Similarly, Titan 2024, Titan 2021 and Kurucz Solar show a concentration of very strong line detections near 6300 \AA, 6900 \AA, 7200 \AA \hspace{0.1mm} and 7600 \AA, precisely where strong visible telluric absorption bands are known to be present \citep{Rudolf2016}, and where the strongest line detections of the telluric calibration observations of Star 2024 are obtained. Using these 2 ``calibrators'' for contamination lines from telluric (Star 2024) and solar (Kurucz Solar) origins, we devised a framework to exclude non-Titan lines from our line selection:

\par \textbf{\textit{Step 1)}} Detected spectral lines in the Titan 2024 and Titan 2021 sets whose central wavelengths fall within the measured FWHM of a Kurucz-Solar line are discarded, being characterized as a ``line of solar origin on Titan's spectrum'' (see figure \ref{fig:lines_spectra}a, shown as red dots, with the Kurucz-Solar lines FWHM shown in red vertical bands). After removing lines characterized as solar, we shall call the resulting line lists ``Non Solar Titan 2024'' and ``Non Solar Titan 2021''. This yielded 5957 ``Non Solar Titan 2024'' lines and 6092 ``Non Solar Titan 2021'' lines (see figure \ref{fig:Detected_Lines_4_steps}a for the Non Solar Titan 2024 lines, and figure \ref{fig:Titan_2021} in the appendix for the Non Solar Titan 2021 lines).

\par After Step 1, we have split the line characterization analysis in two paths: \textbf{Path 1}, which returns a final CH$_4$ feature linelist solely based on the Titan 2024 observations, i.e., without comparing and cross-matching with the Titan 2021 observations. This returns the final RRS-2026 (2 Step) linelist; and \textbf{Path 2}, where there is a cross-match between Titan 2024 and Titan 2021 lines, providing a more stringent linelist than the one in path 1. This returns the final RRS-2026 (3 Step) linelist.

\par \textbf{Path 1, Step 2)} Spectral lines in the Non Solar Titan 2024 line set are compared with the "Star 2024" lines (proxy for telluric absorption lines). We discard all Titan lines which are found within the measured FWHM of a telluric line (in "Star 2024"), being described as a "line of telluric origin on our observation of Titan's spectrum". In figure \ref{fig:lines_spectra}b these lines are shown as blue dots, with the Star 2024 (telluric) lines measured FWHM shown in blue vertical bands. After removing the lines characterized as telluric, we are left with a final linelist of Titan absorption features, which we call "RRS-2026 (2 step) linelist, with a total of 5806 lines, as shown in figure \ref{fig:Detected_Lines_4_steps}b".

\textbf{Path 2, Step 2)} Spectral lines in ``Non Solar Titan 2024'' and ``Non Solar Titan 2021'' have their wavelengths compared  (see figure \ref{fig:Titan_2021} in appendix A). Only the lines appearing in both observations in common wavelengths, i.e., both lines central wavelengths coincide within each others measured FWHM, are kept. Those lines are then assigned the Non Solar Titan 2024 wavelengths. This is a more conservative approach than the one taken in Path 1, retaining only spectral lines observed in Titan's spectrum in two distinct instances, away from any contaminating solar feature. We shall call this line list ``Non Solar Titan 2024 \& 2021'' (see figure \ref{fig:Detected_Lines_4_steps}c), yielding 5034 lines.

\textbf{Path 2, Step 3)} Finally ''Non Solar Titan 2024 \& 2021'' lines are compared to the ``Star 2024'' lines. Similarly to Step 2 or Path 1, we discard all Titan lines found within the measured FWHM of a telluric line (in ''Star 2024''). This more stringent line characterization process yielded a total of 4997 Titan lines, the list of which we call ''RRS-2026 (3 step)'' linelist, as shown in figure \ref{fig:Detected_Lines_4_steps}d. All ''RRS-2026 (3 step)'' lines are also found in ''RRS-2026 (2 step)'', presenting a more conservative approach which requires the identification of these spectral lines in 2 distinct observations of Titan. In table \ref{tab:Linelist_RRS} we present the first lines of our RRS-2026 (2-step) linelist, and whether these lines are included in the more restrictive RRS-2026 (3-steps) linelist. This table corresponds to the first lines of the .txt supplementary to this paper containing the full RRS-2026 linelist.

\par It is worth mentioning that rather than a complete linelist of CH$_4$ in optical wavelengths, this methodology strives for accuracy, following a conservative approached aimed at discarding any line without a Titan origin from this linelist, similarly to what was done in \cite{rianco-Silva2024}. Following the approach of this and other past studies that used Titan's atmosphere as a proxy to study the spectrum of CH$_4$ \citep{Karkoschka1994,Karkoschka2010,Sithajan2025} we associate these detected absorption lines on Titan's spectra to CH$_4$ molecular absorption lines. This is due to the fact that CH$_4$ is the sole molecular species present in Titan's atmosphere known to have absorption bands in the 500 nm to 800 nm wavelength range \citep{Giver1978}. This does not occur in the Solar System giant planets' atmospheres, which, beyond CH$_4$, host significant amounts of gases like NH$_3$ and H$_2$, which also produce meaningful visible absorption bands \citep{Karkoschka1998,Irwin2019}. It is nonetheless important to note that this is an empirical CH$_4$ linelist, and contributions from minor species cannot be categorically ruled out at very high resolution. This is since the minor species known to be present in Titan - particularly more complex hydrocarbons and nitriles - also lack high-resolution linelists in optical wavelengths, as these are even more complex species than CH$_4$. Despite this, it is safe to say that CH$_4$ clearly dominates molecular absorption in the optical regime in Titan and that the vast majority (if not all) of the spectral features identified here correspond to CH$_4$ absorption, as 1) no other low resolution feature in the optical spectrum of Titan is attributable to any other species \citep{Karkoschka2010}, 2) the most abundant second-order hydrocarbon in Titan, C$_2$H$_6$, reaches maximum stratospheric abundances of 20 ppm, 3 orders of magnitude less abundant than CH$_4$ \citep{Coustenis2003}, as also noted by \cite{Sithajan2025}, and 3) the observed distribution of the high resolution spectral lines we identify here neatly matches the low resolution spectrum of CH$_4$.

\begin{table}[ht]
\centering
\caption{First 10 spectral lines from the RRS-2026 (2-steps) linelist. The full table is provided in a machine-readable format as supplementary data to this work. We provide, for each identified spectral line, their measured central wavelength, full width at half maximum (FWHM), relative line depth, and the line depth uncertainty - as well as a yes/no flag regarding whether this line is also included in the more stringent RRS-2026 (3-steps) list.}
\begin{tabular}{ccccc}
\hline
Wavelength (\AA) & FWHM (\AA) & Line Depth & Line Depth Error & In RRS-2026 (3-steps)? \\
\hline
5417.431 & 0.070 & 0.0262 & 0.0058 & Y \\
5417.738 & 0.029 & 0.0165 & 0.0025 & N \\
5418.217 & 0.039 & 0.0219 & 0.0103 & Y \\
5418.895 & 0.033 & 0.0261 & 0.0031 & Y \\
5419.084 & 0.029 & 0.0168 & 0.0025 & N \\
5420.431 & 0.277 & 0.0256 & 0.0031 & Y \\
5421.390 & 0.061 & 0.0232 & 0.0015 & Y \\
5422.963 & 0.285 & 0.0159 & 0.0010 & Y \\
5423.406 & 0.285 & 0.0503 & 0.0121 & Y \\
5424.094 & 0.099 & 0.0332 & 0.0030 & Y \\
... & ... & ... & ... & ... \\
\hline
\end{tabular}
\label{tab:Linelist_RRS}
\end{table}

\begin{figure}%[tbhp]
\centering
\includegraphics[width=0.5\linewidth]{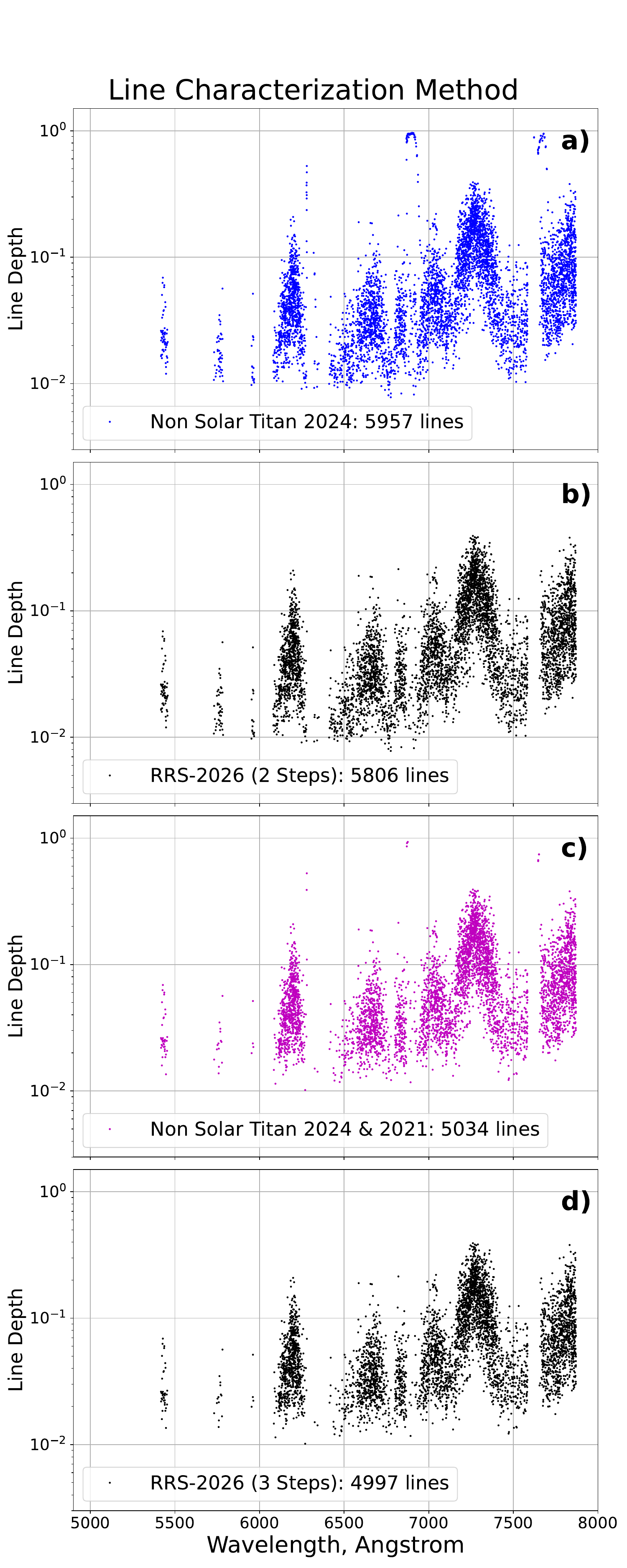}
\caption{Line detections across the distinct stages of the line characterization method: a) ``Non Solar Titan 2024 lines'', resulting from step 1;  b) ``RRS-2026 (2 steps)'' resulting from step 2 of Path 1; c) ``Non Solar Titan 2024 \& 2021'', resulting from step 2 of Path 2 and d) ``RRS-2026 (3 steps)'' resulting from step 3 of Path 2. The two sets of RRS-2026 linelists shown in plots b) and d) are also shown, combined, in Figure \ref{fig:Comparison_Karkoschka}.}
\label{fig:Detected_Lines_4_steps}
\end{figure}

\par It is also important to clarify that the RRS-2026 linelist we extract from Titan's atmosphere in this work is purely observational. Hence, this linelist contains only the detected CH$_4$ line wavelengths and relative intensities - without any transition assignment or temperature dependence, as theory-based linelists produced by groups like ExoMol tend to contain \citep{Tennyson2012}. Still, despite our linelists empirical and incomplete nature, alongside their dependence on the specific conditions of the atmosphere of Titan from which it was extracted, we aim to show that they can be used to produce templates suitable to the application of the HRCCS technique in planetary atmospheres different from Titan's.

\subsection{Optical CH$_4$ Cross-Correlation Functions} 

\par After obtaining the RRS-2026 CH$_4$ linelists, we proceed to use them to produce a spectral template (also known as ``mask'') that can be applied in HRCCS. Following the approach used to produce the widely used MANTIS Network masks for Cross-Correlation analysis \citep{Kitzmann2023}, we use the extracted CH$_4$ lines wavelength and line depths and interpolate them upon the observed spectrum wavelength grid, producing a spectral mask, $M_\lambda$, equal to the line depths at the line's wavelengths and 0 elsewhere.

\par
\begin{figure}
\centering
\includegraphics[width=0.7\linewidth]{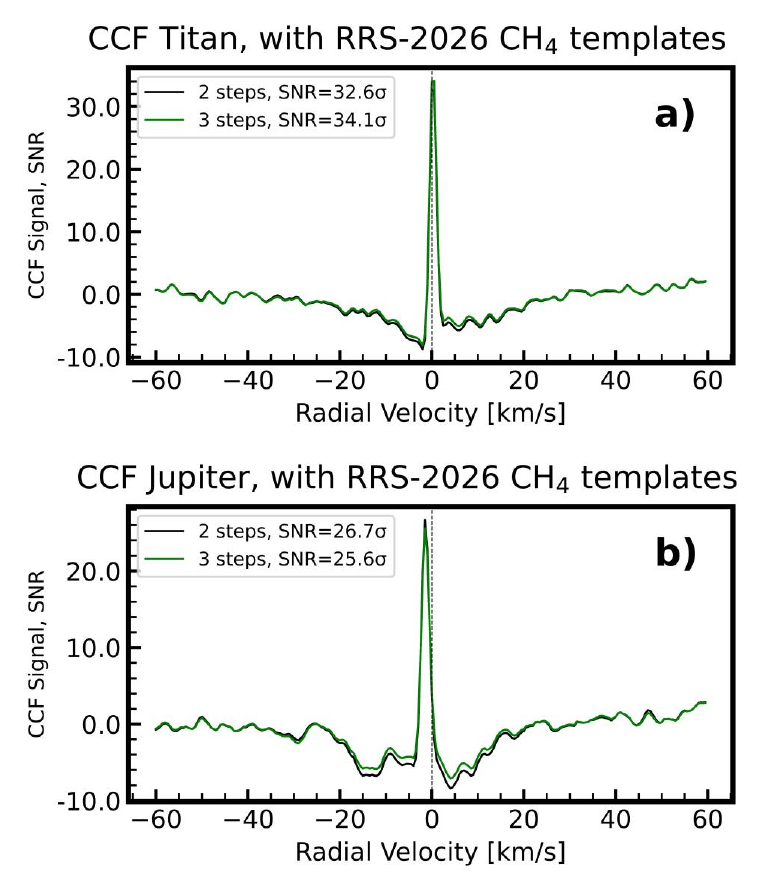}
\caption{Cross-Correlation Functions (CCFs) of VLT-ESPRESSO UHR Spectra of Titan (a) and Jupiter (b), using the CCF templates obtained with the CH$_4$ visible RRS-2026 linelists retrieved in this work. These CCFs yielded detection strengths of CH$_4$ for the Titan VLT-ESPRESSO spectrum of 32.2$\sigma$ for the RRS-2026 (2 steps) linelist and 34.1$\sigma$ for the RRS-2026 (3 steps) linelist. Titan CCFs peak at radial velocities of (0.28 $\pm$ 0.78) km/s and (0.28 $\pm$ 0.80) km/s for the RRS-2026 2-step and 3-step CCFs, respectively. For the VLT-ESPRESSO spectrum of Jupiter, they yielded CH$_4$ detections of 26.7$\sigma$ for the RRS-2026 (2 steps) linelist and 25.6$\sigma$ for the RRS-2026 (3 steps) linelist. Jupiter CCFs peak at radial velocities of (-1.39 $\pm$ 0.98) km/s and (-1.39 $\pm$ 0.99) km/s, for RRS-2026 2-step and 3-step, respectively.}
\label{fig:CCF_RRS}
\end{figure}

 The Cross-Correlation Function, CCF($v$), is then computed by cross-correlating the observed spectrum $S_\lambda$ and the molecular template (or mask), $M_\lambda$ in velocity space. This CCF($v$) yields a strong peak if the template matches the spectrum for a given velocity shift, suggesting a detection whose strength is quantified by how much stronger the detection peak is with respect to the CCF($v$) standard distribution ($\sigma$) away from the peak  \citep{Birkby2018,Snellen2025,Esparza-Borges2023}. Following an approach commonly used when presenting CCF results \citep{Esparza-Borges2023}, we have normalized the CCF with respect to the standard deviation of CCF in the region away from the central peak ($\sigma$), providing the CCF in units of standard deviation, or SNR. For a ``sanity check'' test, we used the RRS-2026 linelist templates to obtain HRCCS CH$_4$ detections from the VLT-ESPRESSO Titan 2024 spectrum - which is shown in figure \ref{fig:CCF_RRS}a. As expected, this yields very clear detection peaks with significances of 32.6$\sigma$ (for the RRS-2026 2 step linelist), and 34.1$\sigma$ (for the RRS-2026 3 step linelist). The peaks are centred at (0.28 $\pm$ 0.78) km/s and (0.28 $\pm$ 0.80) km/s for the RRS-2026 2-step and 3-step CCFs, respectively (with uncertainties corresponding to the CCF HWHM). These are consistent with $v = 0$ km/s as expected due to the fact the RRS-2026 linelist was extracted from this very spectrum.

\par For a more robust test of RRS-2026, it is valuable to apply it to an entirely distinct dataset form the one used to retrieve it as this strong cross-correlation peak observed for the Titan spectrum could have been due to the match between putative spurious template lines and the spurious (i.e., not caused by CH$_4$) features in the Titan 2024 spectrum from which the RRS-2026 linelists were retrieved. Hence, we performed a HRCCS analysis with our CH$_4$ visible template on archived VLT-ESPRESSO UHR observations of Jupiter \citep{Machado2023}, the aforementioned ``Jupiter 2019'' dataset. This enables testing of the applicability of our RRS-2026 linelist to a completely distinct planetary atmosphere, with distinct chemistry, pressure and temperature regimes, and pressure-broadeners (H$_2$/He in Jupiter vs N$_2$ in Titan \citep{Irwin2019}), validating the applicability of RRS-2026 across distinct planetary settings. The resulting CCF is shown in figure \ref{fig:CCF_RRS}b, where a clear HRCCS detection of CH$_4$ in Jupiter's visible spectrum is obtained with RRS-2026 2-steps and 3-steps linelist, yielding CCF detection signals of 26.7$\sigma$ and 25.6$\sigma$, respectively. This is the first ever HRCCS detection of CH$_4$ from visible high-resolution spectra of a planetary atmosphere using a visible high resolution CH$_4$ linelist. %Cool!

\par Unlike what happens with the Titan CCF, the peaks on Jupiter CCF on figure \ref{fig:CCF_RRS}b do not coincide with a null velocity, but are instead centred at velocities of (-1.39 $\pm$ 0.98) km/s and (-1.39 $\pm$ 0.99) km/s, for RRS-2026 2-step and 3-step, respectively. Uncertainty corresponds to the CCF peak's HWHM. This velocity signal is due to Jupiter's radial velocity with respect to the Solar System's Barycenter ($v = -0.60$ km/s at the night of the observation), added to Jupiter's fast rotation, which imprints a Doppler shift on the observed spectrum \citep{Machado2023}. The VLT-ESPRESSO spectrum used in this analysis corresponds to position 36 of the observation as described in \cite{Machado2023}, at a latitude of 10ºS and a longitude of 5,24º West from the sub-solar meridian. Due to Jupiter's rotational velocity of 12.6 km/s, the projected radial velocity of Jupiter's rotation at the small region of Jupiter probed by the VLT-ESPRESSO fiber was of $v_r = 12.6\times \cos(-5.24 ^{\circ}) = -1.15$km/s. This yields an expected total relative velocity with respect to the SS barycenter of -1.75 km/s, which is in agreement with the $v = (-1.4 \pm 1.0)$ km/s measured. 

\section{Discussion: Comparing obtained CH$_4$ visible linelists to past work}
\label{discussion}

\subsection{Comparison with low-resolution cross-sections}

\par This work presents the first high-resolution linelist and HRCCS template of CH$_4$ across the entirety of the visible spectrum, "RRS-2026". Past works have also focused on CH$_4$ spectral features in visible wavelengths, albeit with different approaches and end results, which we compare and discuss in this section.

\par Up to now, the state-of-the-art knowledge of CH$_4$ visible spectra corresponded to several studies by Karkoshcka and co-authors \citep{Karkoschka1994,Karkoschka2010,Karkoschka1998}. These works retrieved low-resolution ($R \sim 200$) CH$_4$ cross sections in visible and near infrared wavelengths obtained by remote and \textit{in situ} observations of the atmospheres of Titan and Solar System gas giants. They outlined the low-resolution structure of CH$_4$ visible absorption bands without explaining the structure that gives rise to these absorption bands \citep{Bourdon2009}. These works complement earlier works such as \cite{Giver1978}, which assigned vibrational overtones to the several visible CH$_4$ bands, and \cite{Hayden-Smith1990}, which described the asymmetry of the 620 nm CH$_4$ band. 

\par
\begin{figure}
    \centering
    \includegraphics[width=0.8\linewidth]{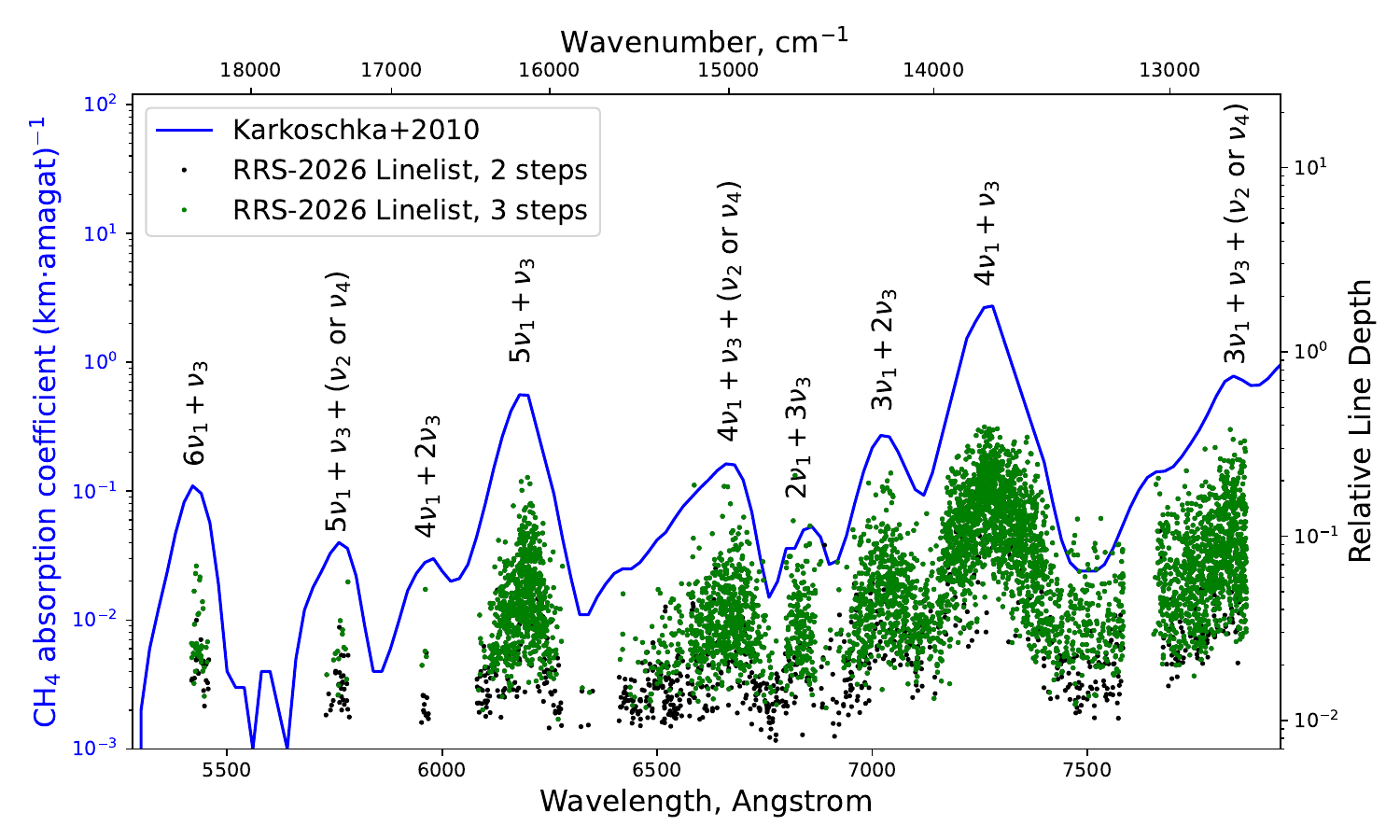}
    \caption{Comparison between this work's RRS-2026 visible CH$_4$ linelists: the more complete RRS-2026 (2 steps), also shown in figure \ref{fig:Detected_Lines_4_steps}b (here in black), which contains the entirety of the more conservative RRS-2026 (3 steps), also shown in figure \ref{fig:Detected_Lines_4_steps}d (here in green), and \cite{Karkoschka2010} visible CH$_4$ absorption cross-section, in blue, with CH$_4$ band assignment following \cite{Giver1978}.}
    \label{fig:Comparison_Karkoschka}
\end{figure}

 In this work we produce a visible high resolution linelist of CH$_4$ absorption lines. We aimed to use this knowledge to explain the broad, low-resolution structure of the visible CH$_4$ cross-sections described above \citep{Karkoschka2010}. We compare our detected CH$_4$ line depth distribution as a function of wavelength with \cite{Karkoschka2010} CH$_4$ absorption cross-section, as shown in figure \ref{fig:Comparison_Karkoschka}, highlighting the matching distribution of our detected CH$_4$ lines and the previously known CH$_4$ absorption bands, in low spectral resolution.

\par 

\subsection{Comparison with near-infrared high-resolution linelists}

\par As previously mentioned, the high-resolution structure of CH$_4$ visible absorption bands is largely unknown and challenging to study, both from theoretical and experimental means \citep{Bourdon2009}. A recent study by \cite{Campargue2023} aimed at pushing the boundaries of high-resolution CH$_4$ linelist characterization to near-infrared wavelengths below 1 $\mu$m. For this, they explored the near-infrared CH$_4$ bands down to 720 nm. Hence, there is an overlap with 2 of the CH$_4$ bands characterized in our study at the red end of VLT-ESPRESSO spectrum - the 4$\nu_1$ + $\nu_3$ (centred at 727 nm) and the 3$\nu_1$ + $\nu_3$ + ($\nu_2$ or $\nu_4$)  (centred at 790 nm) overtone bands, following \cite{Giver1978} overtone band assignment.

\par
\begin{figure}
    \centering
    \includegraphics[width=0.7\linewidth]{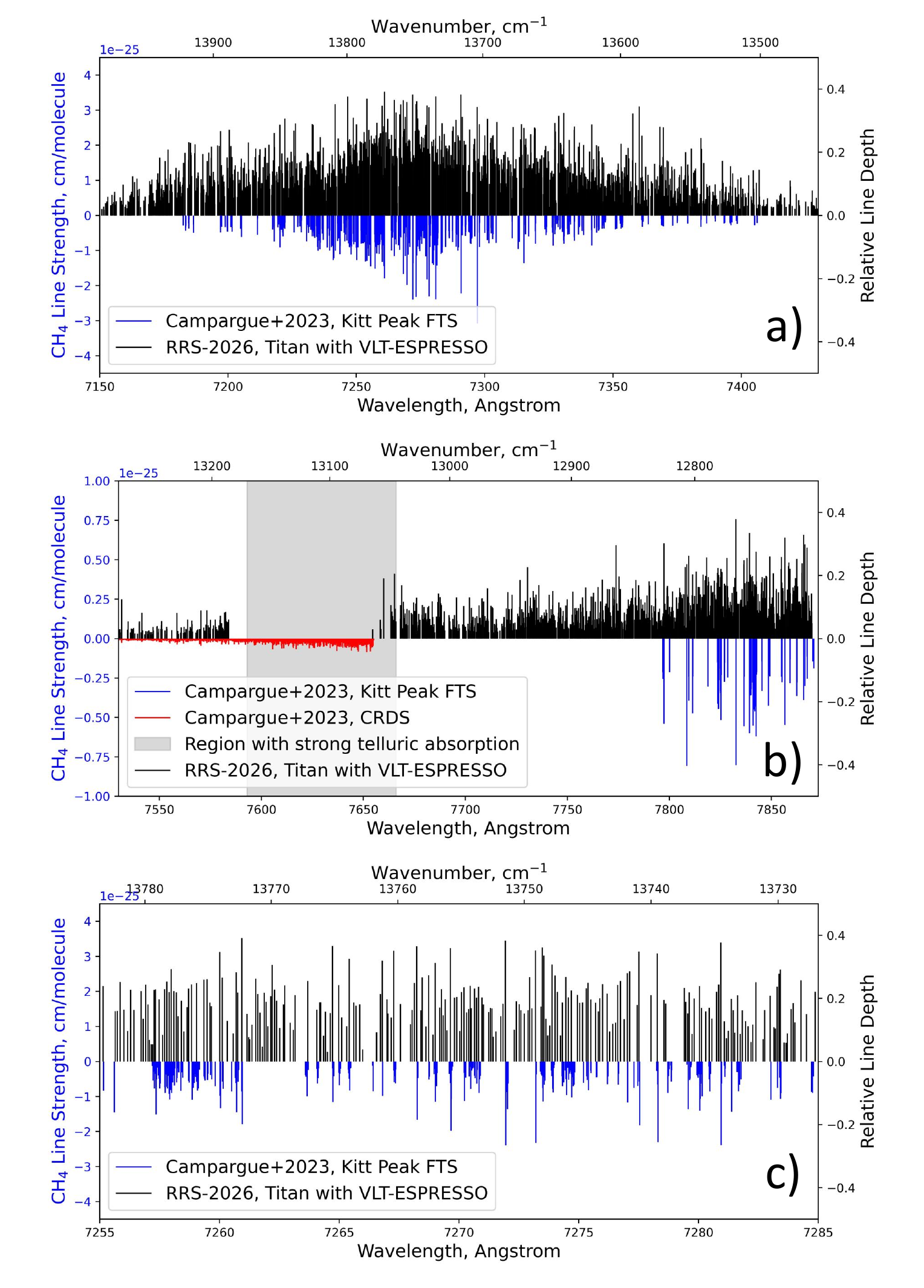}
    \caption{Comparison between the RRS-2026 CH$_4$ (2-step) linelist and the CH$_4$ linelists described in \cite{Campargue2023}, overlapping only in 2 CH$_4$ visible absorption bands above 715 nm, shown in figues a) and b). Figure c) shows a zoom into the central, densest region of the 727nm CH$_4$ band, allowing a comparison between the finer structure of both linelists.}
    \label{Campargue_Comparison}
\end{figure}

 In figure \ref{Campargue_Comparison} we showcase the comparison of the RRS-2026 linelist with the linelists described in \cite{Campargue2023}. These result from a discontinuous coverage of the CH$_4$ optical and NIR spectrum by the Kitt Peak Spectrograph (KTS, as shown in blue) as well a small spectral section covering a region between stronger CH$_4$ bands, observed with Cavity Ring Down Spectroscopy (CRDS, shown in red) \citep{Campargue2023}.

\par A general similarity in terms of band structure and distribution of spectral features is observed between the Kitt Peak FTS linelist and RRS-2026, particularly for the strongest spectral features - suggesting a higher sensitivity of our method to weaker features in these probed spectral regions. On the contrary, the CRDS measurement (also focused in a narrow wavelength range) appears to detect a large number of weaker CH$_4$ features which are unmatched in our work. This is can be attributed to a larger sensitivity to weaker CH$_4$ features by the CRDS, but also due to the specific wavelength range it covers; in the VLT-ESPRESSO observations, this spectral region overlaps with a strong telluric absorption feature which significantly hampers spectral detections in other planetary atmospheres from a ground-based observatory \citep{Rudolf2016}. In figure \ref{Campargue_Comparison}c, we present a zoom on the central, densest region of the 727nm CH$_4$ band, enabling a comparison between the high resolution structure of both measurements. We observe a denser array of spectral features in the RRS-2026 linelist (possibly indicating higher sensitivity from our natural spectroscopy laboratory approach to weaker lines), while features in \cite{Campargue2023} appear to be more concentrated, possibly a result of the higher spectral resolution of the Kitt Peak FTS instrumentation. Notably, at the individual line level, the strongest features identified in \cite{Campargue2023} also match the strongest features in RRS-2026.

\par Most crucially, the CH$_4$ line detections presented in \cite{Campargue2023} cover only narrow sections of the CH$_4$ visible bands. This contrasts with the wide wavelength coverage of our linelist, covering the entirety of the visible spectrum which makes it more suitable to generate HRCCS templates, as wider wavelength coverage has been shown to play a major role in HRCCS detction efficiency \citep{Snellen2025,Giacobbe2021}. 

\subsection{Comparison with past high spectral resolution observations of Titan}

As discussed above, Titan's atmosphere has been used in the past as a natural spectroscopy laboratory for the study of methane (CH$_4$) visible bands, both at low \citep{Karkoschka1994,Karkoschka2010} and high spectral resolutions \citep{rianco-Silva2024,Sithajan2025}. The first study aimed at identifying high-resolution absorption features on Titan's visible spectrum, \cite{rianco-Silva2024}, identified 97 CH$_4$ absorption features in Titan's visible spectrum between 500 and 619 nm, using archived VLT-UVES observations (R $\sim$ 100000). Despite laying the ground-work for a survey of CH$_4$ features on Titan's visible spectrum, the limited wavelength range in these VLT-UVES (shortward of 619 nm) prevented a complete survey of Titan's visible CH$_4$ spectrum - motivating this work with VLT-ESPRESSO, which covers the entirety of Titan's visible spectrum at twice the spectral resolving power reached by VLT-UVES.

\par
\begin{figure}
    \centering
    \includegraphics[width=0.8\linewidth]{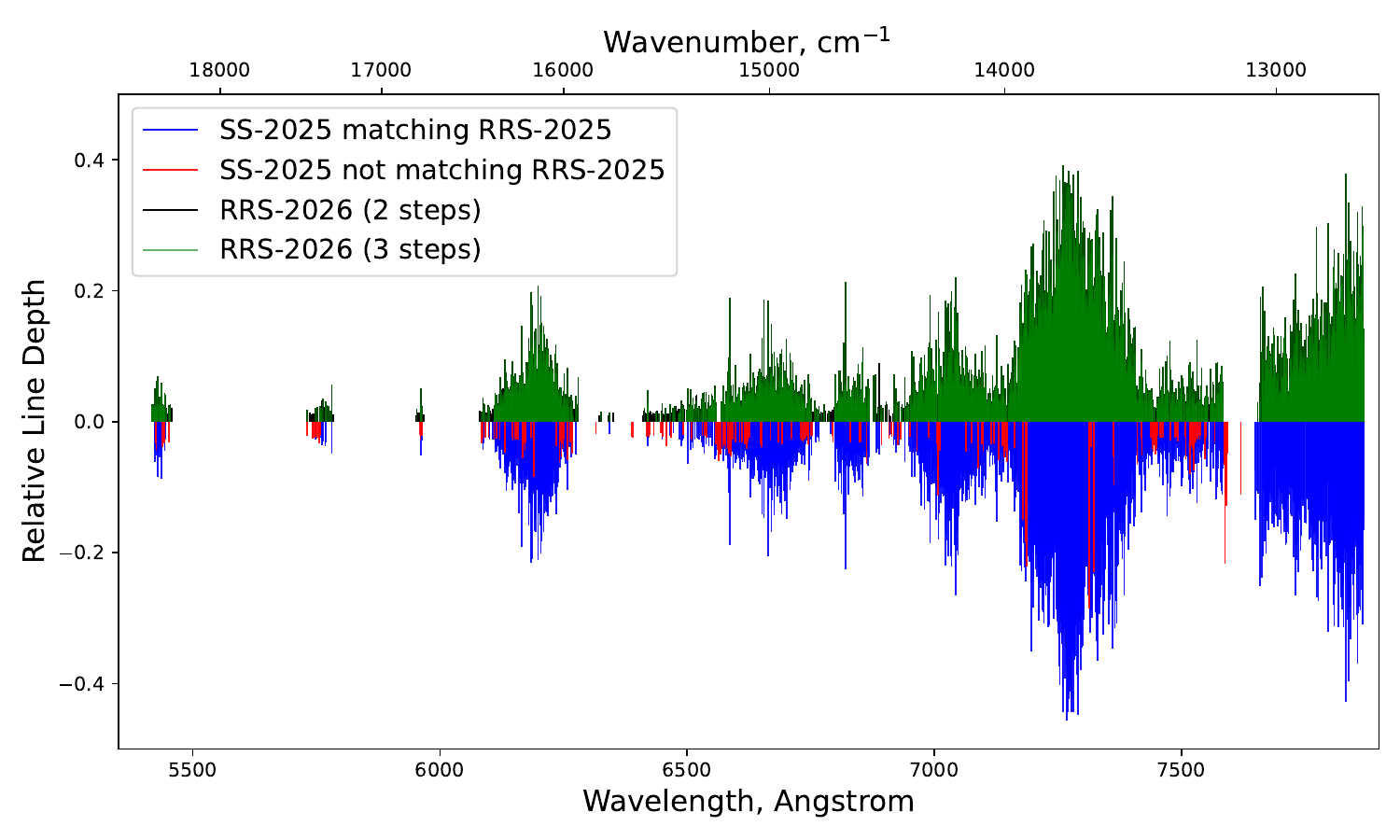}
    \caption{Comparison between this work's CH$_4$ visible high resolution linelist, RRS-2026 (2-steps, in black, and RRS-2026 (3-steps), in green, with \cite{Sithajan2025} Titan spectral features, associated to CH$_4$ absorption features (SS-2025). In blue we identify the SS-2025 features that match in wavelength spectral lines in RRS-2026 (2 step, the most comprehensive line set), whereas in red are shown the spectral features in SS-2025 that are not matched by any feature in RRS-2026. Some of these mismatches could be associated to erroneous identification of spectral features of telluric or solar origin into SS-2025.}
    \label{fig:Comparison_SS2025}
\end{figure}

 Another recent work, by \cite{Sithajan2025}, used archived VLT-ESPRESSO observations of Titan (the same dataset used to extract the Titan 2021 lines used in this work) to extract an “intrinsic Titan spectrum''. This was done by fitting external contributions to the high-resolution spectrum of Titan (such as solar or telluric features) and dividing these out of the spectrum - in theory returning a high-resolution CH$_4$ absorption spectrum from Titan's atmosphere. In their work, the authors use this intrinsic Titan spectrum to extract 6195 absorption features within the 5400–7870 \AA \vspace{0.1mm} range, potentially associated with CH$_4$ absorption. It is also mentioned that 1207 of these features are found near strong solar or telluric absorption features, which may hamper their reliability as CH$_4$ detections. \cite{Sithajan2025} provide this intrinsic VLT-ESPRESSO Titan spectrum, as well as the list of identified spectral features and a mask allowing to exclude the spectral features identified as near contaminating solar or telluric features. We use this set of 6195 Titan features obtained by \cite{Sithajan2025}, mentioned henceforth as ``SS-2025'' - to compare with this work's RRS-2026 CH$_4$ linelist. This comparison is shown in figure \ref{fig:Comparison_SS2025}, where we show in blue the spectral features in SS-2025 that are found to match  spectral features in RRS-2026 (2-step) linelist within each RRS-2026 line's FWHM - and in red the features in SS-2025 that do not match any RRS-2026 feature.

\par Out of all the spectral features in SS-2025, 5790 of these are found to match features identified in RRS-2026 - which means that these features were attributable to CH$_4$ by two different methods, suggesting their reliability as very likely CH$_4$ absorption lines. We note that, despite the efforts taken to remove non-CH$_4$ contributions to the SS-2025 Titan intrinsic spectrum by \cite{Sithajan2025}, we verified that there remain features that are not matched by the features observed in RRS-2026, but instead can be attributable to solar or telluric lines (405 lines in the full SS-2025 line set and 315 lines in the masked, higher confidence, SS-2025 lines). This is likely caused by an incomplete removal of contaminating non-Titan features from Titan's spectrum due to the challenging fit to these deep solar or telluric features. These include absorption features at 760 nm (which we associate to strong telluric absorption features, see figure \ref{fig:Comparison_SS2025}). %In our independent analysis of SS-2025 we were thus able to identify as non-Titan features (due to their proximity with solar or telluric features) 1055 features, which alongside their non-match to RRS-2026 features indicates these are a lower limit of possible contaminating telluric or solar features present in SS-2025.

\par The key difference between our approach and the one followed by \cite{Sithajan2025} is how contaminating sources to Titan's intrinsic spectrum are considered. While we have used other observations of a telluric calibrator star and a solar spectrum to flag features of solar or telluric origin on Titan's spectrum, allowing to exclude them from our extracted linelist, in \cite{Sithajan2025} a fit to these telluric and solar contributions on Titan's spectrum was applied - with the residuals being interpreted as being originated in Titan. This leaves the work of \cite{Sithajan2025} vulnerable to imperfect fits to solar and telluric contributions in the observed spectrum, that can enable spurious features to show up in the residual, which could be mistakenly interpreted as CH$_4$ features of Titan's origins. Despite our dedicated observations' larger SNR, our conservative line identification strived to rule out possible false CH$_4$ features, as HRCCS studies are highly sensitive to the quality of spectral templates \citep{Birkby2018,Yurchenko2024}. This led us to obtain a similar number of identified spectral features as \citep{Sithajan2025}, despite applying the same SNR threshold for line detection (a line depth 5 times larger than the local spectral SNR).

\par
\begin{figure}
    \centering
    \includegraphics[width=0.7\linewidth]{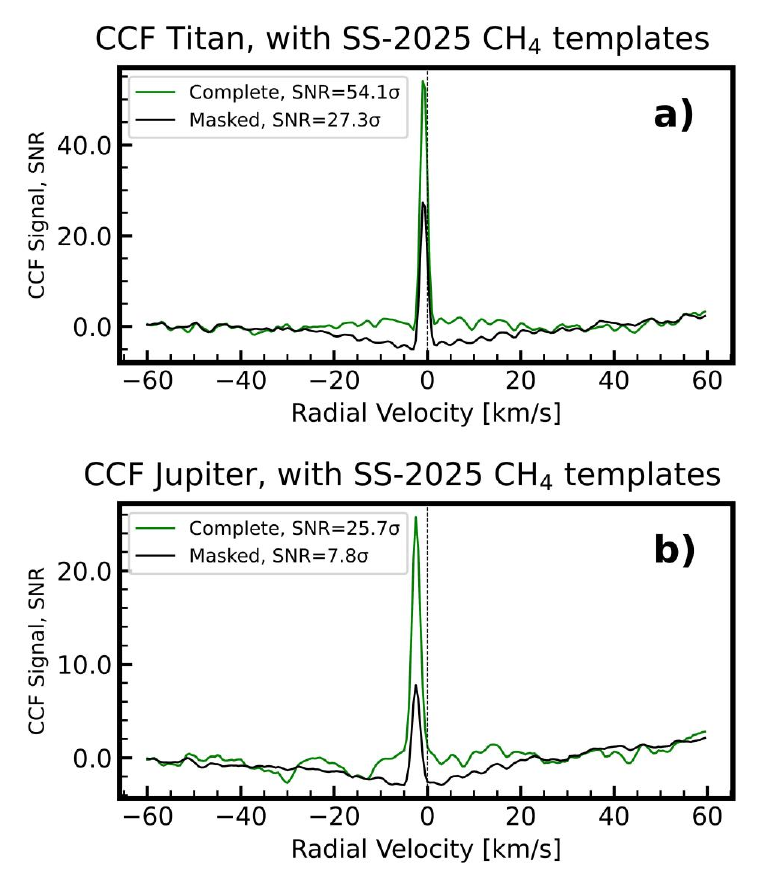}
    \caption{Cross-Correlation Functions (CCFs) of VLT-ESPRESSO UHR Spectra of Titan (a) and Jupiter (b), using the CCF templates obtained with the CH$_4$ visible SS-2025 linelists retrieved in \cite{Sithajan2025} (complete or masked for contaminating sources). These CCFs yielded detection strengths of CH$_4$ for the Titan VLT-ESPRESSO spectrum of 54.1$\sigma$ for the SS-2025 (complete) linelist and 27.3$\sigma$ for the  SS-2025 (masked, uncontaminated) linelist with both CCFs peaking at radial velocities of (-0.77 $\pm$ 0.83) km/s. For the VLT-ESPRESSO spectrum of Jupiter, it yielded CH$_4$ detections of 25.7$\sigma$ for the SS-2025 (complete) linelist and 7.8$\sigma$ for the SS-2025 (masked, uncontaminated) linelist. Central peaks velocities from SS-2025 Jupiter CCFs were (-2.43 $\pm$ 1.08) km/s and (-2.43 $\pm$ 0.91) km/s for the complete and masked SS-2025 linelists, respectively.}
    \label{fig:CCF_SS2025}
\end{figure}

 We test HRCCS performance with SS-2025 by using both the unmasked, complete SS-2025 set (with 6195 lines) and the masked, higher-confidence SS-2025 line set (with 4988 lines) to produce HRCCS templates (following the approach described in the previous section) and extract HRCCS detections of CH$_4$ in the Titan and Jupiter datasets - as shown in figures \ref{fig:CCF_SS2025}a) and \ref{fig:CCF_SS2025}b), respectively. While the SS-2025 template allowed HRCCS detections of CH$_4$ from VLT-ESPRESSO observations of both Titan and Jupiter, these detections yielded distinct detections strengths compared to HRCCS detections conducted with the RRS-2026 template, and within SS-2025 templates, the complete and masked templates also yielded very different results. For Titan, the SS-2025 complete template yielded a significantly stronger cross-correlation signal with the VLT-ESPRESSO data than the RRS-2026 linelists (54.1$\sigma$) - while the template where lines flagged as possibly contaminated were masked out yielded a lower detection strength (27.3$\sigma$), also lower than the ones obtained by RRS-2026. This does not necessarily imply that it is a better CH$_4$ linelist; one criterion for assessing the quality of a molecular linelist obtained through a natural laboratory, such as these, is to test the linelist on a distinct target to ensure the linelists can be unambiguously assigned to a molecule rather than the atmosphere. In the case of SS-2025, the reduced SNR of the CCF resulting from the masked linelist could be due to the fact that when masking these lines out, good CH$_4$ lines are actually being removed from the template, decreasing the detection significance. However, a more problematic possibility could be the fact that the complete SS-2025 template includes spurious, non-CH$_4$ lines, that reflect contaminating sources observed on Titan's spectrum (i.e. solar features that are picked up by the CCF, spuriously increasing the detection signal). 
\par We measured central peaks velocities of -0.77 $\pm$ 0.83 km/s for both complete and masked SS-2025 templates, which are still consistent with null velocity at the edge of the measured uncertainty - but further away from v = 0 km/s when compared to RRS-2026, despite both linelists being computed at vacuum wavelengths. This could reflect the fact that, unlike for RRS-2026, the Titan 2024 dataset was not used to extract the SS-2025 lines. This could also reflect the lower SNR of the Titan 2021 ESPRESSO dataset used to extract SS-2025 when compared to the Titan 2024 dataset we used to extract RRS-2026. As for the application of SS-2025 to the Jupiter VLT-ESPRESSO spectrum, CCF detection strengths are significantly lower - a stark contrast in magnitude which we did not observe for RRS-2026. Indeed, even the strongest detection yielded by SS-2025 for Jupiter (obtained by the complete SS-2025 template) is lower in magnitude than the one obtained by our linelist RRS-2026 (2 steps). Central peaks velocities from SS-2025 Jupiter CCFs were (-2.43 $\pm$ 1.08) km/s and (-2.43 $\pm$ 0.91) km/s for the complete and masked SS-2025 linelists, respectively, both consistent with the expected radial velocity signal stemming from Jupiter's radial velocity relative to Earth and its rotation signal (-1.75 km/s). Again, despite both templates being consistent within their errorbars with the expected velocity, the lower absolute deviation with respect to the true value of the velocity extracted with RRS-2026 strengthens our recommendation of the usage of RRS-2026 as a linelist for CH$_4$ in the optical. We hypothesize that this is a consequence of the higher SNR of the Titan 2024 dataset used in this work.

\par This applicability to two distinct planetary atmospheres (at distinct systematic radial velocities that allow to exclude solar or telluric contamination as a source of CCF detection signals) points to the RRS-2026 (2-steps) linelist as the preferred low-temperature spectral template for CH$_4$ HRCCS: a larger sample of uncontaminated lines retrieved at higher SNR result in stronger HRCCS detections. The extraction procedures we find the most appropriate to obtain such an uncontaminated sample of new spectral lines from a natural spectroscopy laboratory such as this is the conservative approach we took. This required the characterization of all observed lines origins, flagging and masking out possible contaminating lines, rather than fitting for contaminating contributions which risks leaving behind contaminating residuals. Hence, we recommend RRS-2026 (2 steps) as the state-of-the-art CH$_4$ linelist for future HRCCS searches for CH$_4$ with optical high-resolution spectrographs targetting cold ($T <$ 200K) exoplanet atmospheres. While this linelist might also be used in HRCCS searches of CH$_4$ in warmer exoplanet atmospheres, it is worth taking into account that the empirical nature of the RRS-2026 linelist implies that it reflects the temperature conditions of Titan's atmosphere, where 70K $< T <$ 200K \citep{Lunine2008}. This linelist's empirical nature also prevents an assignment of upper or lower energy levels associated with each molecular transition responsible for each line. This also prevents modelling of CH$_4$ optical spectrum at higher temperatures \citep{Yurchenko2024,Kefala2024} and hence we recommend caution when applying this linelist for higher temperature environments, as both average opacity and relative band intensity will be affected \citep{sousa2015exomol, neale1996spectroscopic}. 

\section{Conclusion} High Resolution Cross-Correlation Spectroscopy (HRCCS) is a crucial technique that has enabled significant advances in the study of exoplanet atmospheres, allowing the detection of many atomic and molecular species by combining a large number of weak features hidden in the noise of exoplanet spectra through a cross-correlation with a species-specific spectral template \citep{Snellen2010,Birkby2018,Snellen2025,Giacobbe2021}. The quality of linelists or molecular cross-sections used to produce HRCCS templates is shown to significantly affect the detectability of chemical species in a planetary atmosphere \citep{Yurchenko2025,Brogi2019,Gandhi2020}, as incorrect opacities degrade the template–observed spectrum match, decreasing or even preventing a HRCCS detection. 

\par Methane (CH$_4$) is a molecule of particular interest in the context of exoplanet atmospheres \citep{Fortney2020}, particularly for sub-jovian worlds as temperate sub-Neptunes \citep{Madhusudhan2025}, Titan-like worlds \citep{Lunine2008,Hayes2018}, and as a biosignature in rocky worlds \citep{Thompson2022}. However, due to the lack of a visible high-resolution CH$_4$ spectral linelist \citep{Campargue2023,Yurchenko2024,Kefala2024}, the search for this key molecule using HRCCS has been limited to infrared wavelengths; challenging to observe from the ground-based high-resolution facilities due to telluric contamination \citep{Rudolf2016}. The only opacities available for the CH$_4$ visible spectrum were retrieved from empirical measurements, in low spectral resolution \citep{Karkoschka2010,Karkoschka1998}, recovering broad CH$_4$ bands but not the individual lines required to produce templates for HRCCS.

\par In this work we used the highest spectral resolution observations of Titan's atmosphere to date in the optical, where the sole molecular absorber is CH$_4$ \citep{Karkoschka2010}, to extract a CH$_4$ high resolution visible linelist. To identify spectral lines on Titan's backscattered visible spectrum based on their telluric, solar or Titan origins, we used original VLT-ESPRESSO observations (``Titan 2024'' dataset), alongside older, archived VLT-ESPRESSO observations of Titan (``Titan 2021'' dataset), VLT-ESPRESSO observations of a star for telluric line calibration (``Star 2024'' dataset), and the Kurucz (2006) high resolution solar spectral library \citep{Kurucz2006}. The lines found to originate from absorption in Titan's atmosphere were assigned to CH$_4$ absorption, enabling the production of the RRS-2026 CH$_4$ low-temperature, visible, high-resolution linelist with 5806 identified lines spanning visible wavelengths from 530 nm to 788 nm. Past attempts to measure the visible spectrum of CH$_4$, albeit only probing a small section of the visible spectrum \citep{Campargue2023}, retrieved at low spectral resolution \citep{Karkoschka2010}, or providing an incomplete removal of contaminating sources to Titan's spectrum \citep{Sithajan2025}, are in agreement with RRS-2026, the linelist newly presented in this work.

\par The RRS-2026 linelist was used to build the first CH$_4$ visible, high-resolution HRCCS template, that was then used to perform a cross-correlation analysis with VLT-ESPRESSO observations of Titan and Jupiter. This enabled the first ever HRCCS detection of CH$_4$ in a planetary atmosphere using a visible cross-correlation template - showcasing its applicability in distinct atmospheric conditions (e.g., Jupiter) to those used to extract our linelist. Crucially, this work paves the way for the search for CH$_4$ in exoplanet atmospheres using HRCCS in a previously very challenging and uncharacterised wavelength regime - the optical - opening a new avenue for the search for this key chemical species in the atmospheres of increasingly smaller and cooler exoplanets, as future facilities such as ELT/ANDES are expected to come online in the coming years.

%%%%%%%%%%%%%%%%%%%%%%%%%%%%%%%%%%%%%%%%%%%%%%%%%%
\section{Data Availability}

We present as an online readme file the retrieved RRS-2026 linelist (in the same format as the inline Table 1). We further provide .txt files containing the lists of detected lines in each spectral dataset shown in figure \ref{fig:Detected_Lines_4_targets} (Titan 2024, Titan 2021, Star 2024, Kurucz Solar) as well as the 3 intermediate steps shown in figures \ref{fig:Detected_Lines_4_steps}a and c and in figure \ref{fig:Titan_2021} (Non-Solar Titan 2024, Non-Solar Titan 2024 \& 2021 and Non-Solar Titan 2021, respectively).

\section{Acknowledgments}

This work was supported by Fundação para a Ciência e Tecnologia (FCT) of reference PTDC/FIS-AST/29942/2017, through national funds and by FEDER through COMPETE 2020 of reference POCI-01-0145-FEDER-007672, and through the research grants UIDB/04434/2020, UIDP/04434/2020 and UID/04434/2025. RRS acknowledges funding through the FCT fellowship grant 2024.02527.BD. CSS acknowledges funding through the FCT grant (DOI 10.54499/2022.06872.CEECIND/CP1721/CT0002). This study was based on original observations collected at the European Organisation for Astronomical Research in the Southern Hemisphere under ESO programmes 0103.C-0203(A) and 114.277N (P.I: Machado). This study also used, for further comparison with our original data, the publicly available, archived observations of Titan from the ESO program 106.218L.001 (P.I: Martin Turbet) which we acknowledge here. % Display the acknowledgments section

\begin{contribution}
%%This section gives authors the space to recognize author contributions. The text inside this environment is NOT counted towards the total word quanta. At a minimum, manuscripts are expected to include this text:

RRS and PM performed the observational proposal submitted to ESO and obtained the observational data from ESO. RRS performed the data reduction and analysis and paper writing. RRS, PM, CSS, SY, GT have performed the project conceptualization and manuscript reviewing and editing.

%% But authors are expected to provide more specific details, e.g. 
%%
%%SC was responsible for writing and submitting the manuscript.
%%WWM came up with the initial research concept and edited the manuscript.
%%OTS obtained the funding and edited the manuscript.
%%EBF provided the formal analysis and validation. He also edited the manuscript.
%%GEH Supervised the undergraduates, wrote the software and administers the project github and Zenodo repositories.
%%
%% Authors can use the Contributor Role Taxonomy (CRediT) at
%% for ideas on how write a good statement tailored to their needs.

\end{contribution}

\facilities{VLT-ESPRESSO}

%\software{astropy \citep{2013A&A...558A..33A,2018AJ....156..123A,2022ApJ...935..167A},  
\appendix

\section{Titan 2021 detected, non-solar lines}

\begin{figure}[h]
    \centering
    \includegraphics[width=0.8\linewidth]{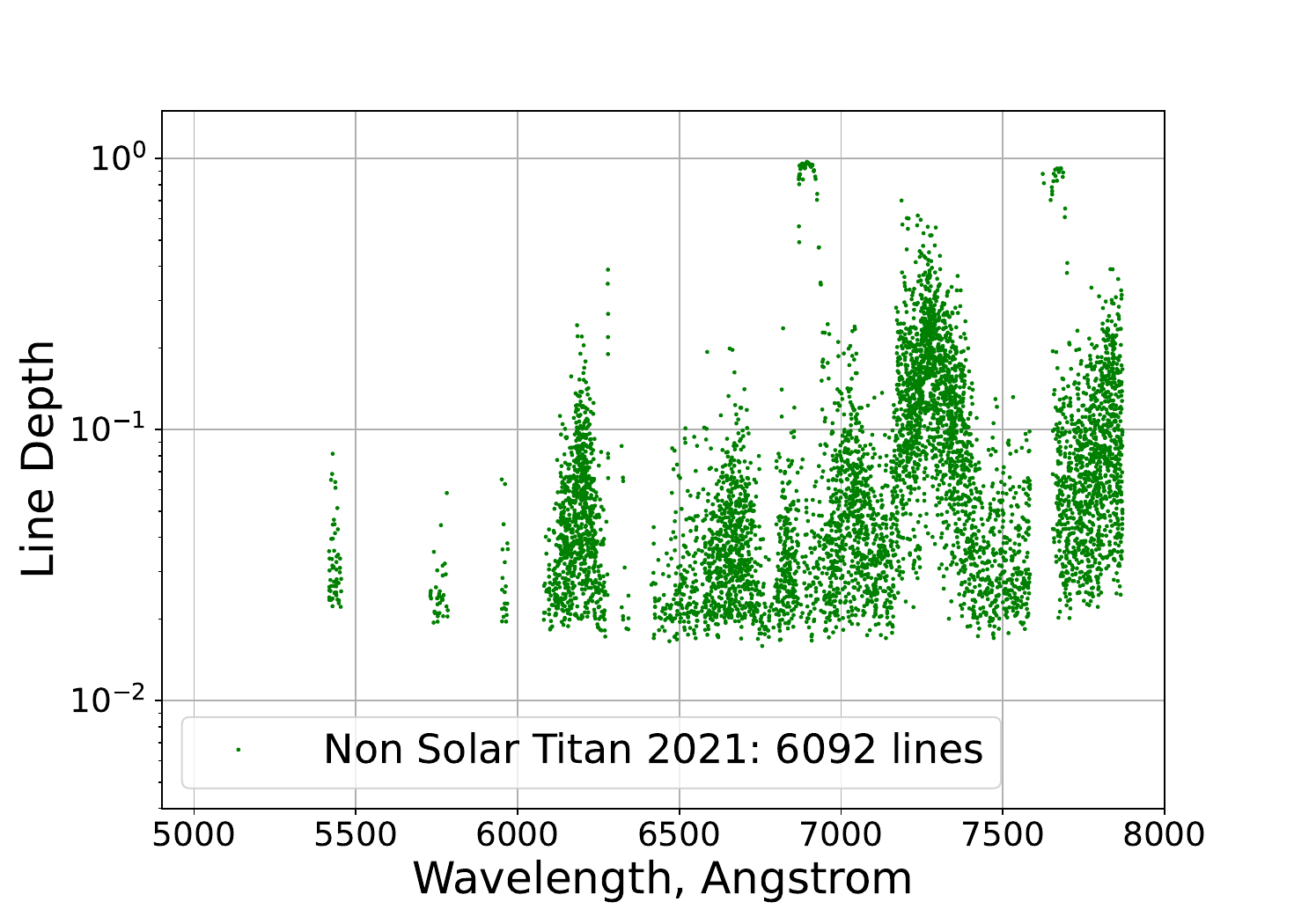}
    \caption{“Non Solar Titan 2021 lines”, resulting from step 1, equivalent to Figure \ref{fig:Detected_Lines_4_steps}a for Titan 2021 lines.}
    \label{fig:Titan_2021}
\end{figure}


%% For this sample we use BibTeX plus aasjournalv7.bst to generate the

\bibliography{sample701}{}
\bibliographystyle{aasjournalv7}



\end{document}